\documentclass[aps,pra,footinbib,showpacs,twocolumn,showkeys,amssymb,amsmath,superscriptaddress]{revtex4-1}

\pdfoutput=1
\usepackage{amsmath}
\usepackage{amsfonts}
\usepackage{braket}
\usepackage{hyperref}
\usepackage[usenames]{color}
\usepackage{natbib}
\usepackage{stackengine}
\usepackage{verbatim}
\usepackage{graphicx}

\begin{document}
\title{Quantum simulation of long range $XY$ quantum spin glass with strong area-law violation using trapped ions}

\author{Nilanjan Roy}
\affiliation{Department of Physics, Indian Institute of Science Education and Research, Bhopal, India}
\author{Auditya Sharma}
\affiliation{Department of Physics, Indian Institute of Science Education and Research, Bhopal, India}
\author{Rick Mukherjee}
\affiliation{Department of Physics, Indian Institute of Science Education and Research, Bhopal, India}
\affiliation{Department of Physics, Imperial College London, SW7 2AZ, London, UK}

\date{\today}

\begin{abstract}
Ground states of local Hamiltonians are known to obey the entanglement entropy area law. While area law violation of a mild kind (logarithmic) is commonly encountered, strong area-law violation (more than logarithmic) is rare.  In this paper, we study the long range quantum spin glass in one dimension whose couplings are disordered and fall off with distance as a power-law. We show that this system exhibits more than logarithmic area law violation in its ground state. Strikingly this feature is found to be true even in the short-range regime in sharp contrast to the spinless long range disordered fermionic model. This necessitates the study of large systems for the quantum $XY$ spin glass model which is challenging since these numerical methods depend on the validity of the area law. This situation lends itself naturally to the exploration of a quantum simulation approach. We present a proof-of-principle implementation of this non-trivially interacting spin model using trapped ions and provide a detailed study of experimentally realistic parameters.
\end{abstract}

\maketitle
\section{Introduction}
\begin{figure}[t!]
	\includegraphics[width=1.0\columnwidth]{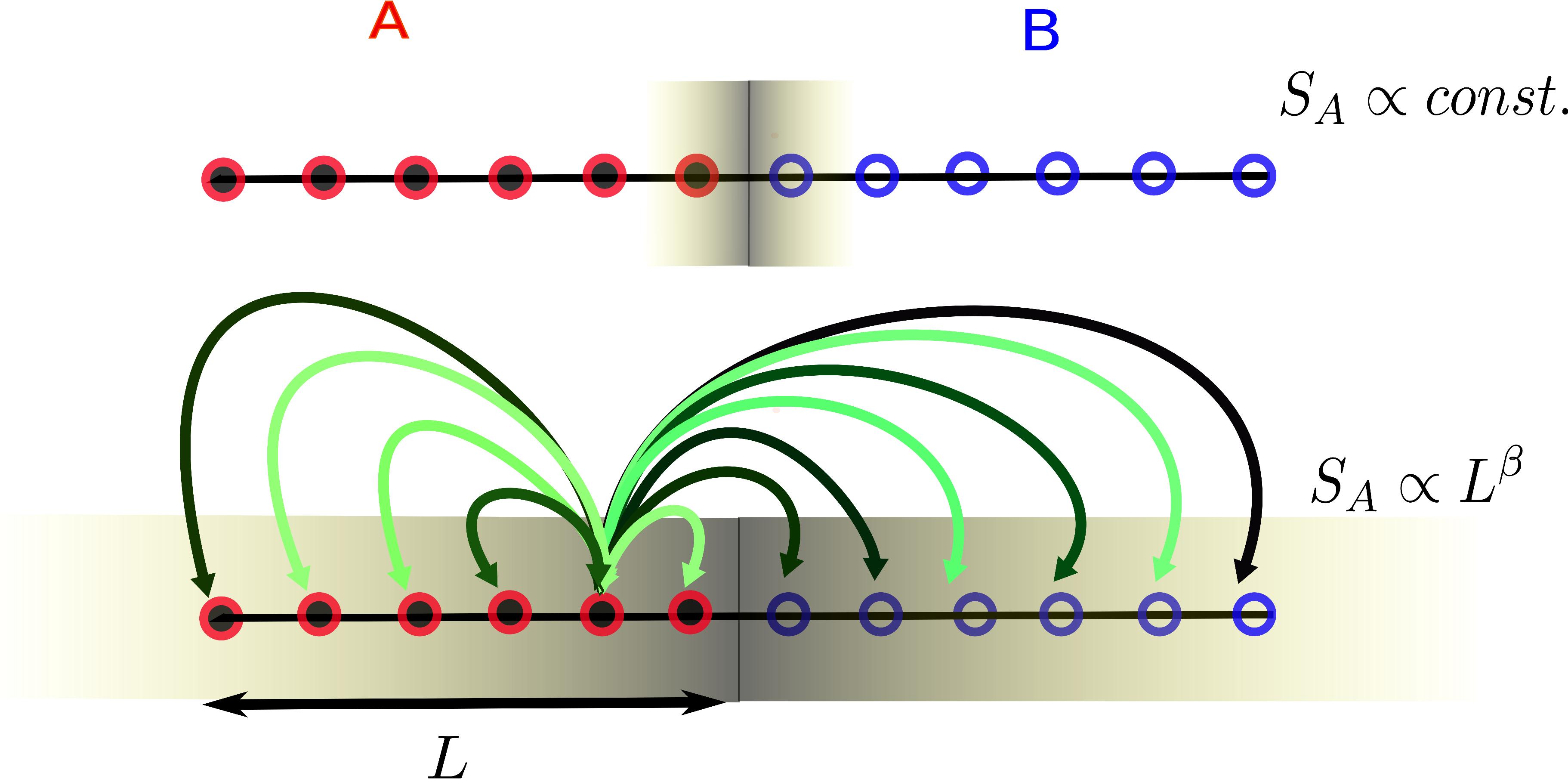}
	\caption{Depicts the violation of area-law of entanglement between two subsystems A (filled red circle) and B (empty blue circle) for any one dimensional quantum system. The area-law of entanglement is usually valid when quantum correlations between A and B (represented by colored shading) exist only at the boundary and exponentially vanish beyond it. However when these correlations sustain even beyond the boundary between A and B, then one obtains violation of the area-law. In this work, we find that for disordered long range spin couplings, represented by the dark(light) green arrows corresponding to randomly assigned positive (negative) values, one obtains strong power-law violation of the area-law of entanglement.}
	\label{setupfig1}
\end{figure}

Entanglement in quantum systems has become one of the most
studied subjects in physics~\cite{horodecki2009quantum,amico2008entanglement,eentropy_calabrese,gorshkov}. The scaling of entanglement entropy with the spatial size of the
subsystem has proved to be a useful diagnostic to characterize
different phases across quantum phase transitions~\cite{eisert2010,laflorencie2016quantum}.  Many-particle
ground state wavefunctions of local Hamiltonians adhere to the
``area-law'' of entanglement\cite{hastings2007area,eisert2010,laflorencie2016quantum,srednicki},
according to which, in a subsystem of volume $L^d$ in a
$d$-dimensional system, entanglement entropy scales as $S_A\propto
L^{d-1}$, since only the degrees of freedom near the boundary of the
two subsystems contribute to the entanglement entropy (Fig.~\ref{setupfig1}). The well-known logarithmic violation of the area law ($S_A\propto L^{d-1}
\ln L$) has been reported in a variety of translationally invariant
models such as gapless systems (e.g. quantum critical
point)~\cite{vidal2003,latorre2003,korepin2004universality,Jin2004},
free fermionic (hopping)
models~\cite{wolf2006violation,klich,swingle,gori2015explicit},
bosonic systems~\cite{lai2013violation,ding2009} and interacting spin
chains~\cite{peschel2004,eisert2006,korepin2005,popkov2005,ding2008,riera2009,Lewenstein,daley,frerot_2017,vodola_2015}.  Quantum dynamics of correlations in the long range spin chains with $\ln L$-kind violations are known to show many exotic features~\cite{liu2018,safavi2018,Smith1,zoller2014},  

Strong area-law violations indicate very strong correlations and highly-entangled ground sates could be useful for quantum computation technologies. However, ground states showing stronger-than-logarithmic violations of the area law ($S_A\propto L^{\beta}$), particularly in systems that preserve translational symmetry are rare~\cite{gori2015explicit,volumelaw_SDSarma}. There are also exotic special constructions like the Motzkin spin chain~\cite{bravyi2012criticality,movassagh2017entanglement} where the Hamiltonian is theoretically engineered so as to make a pre-designed area-law-violating state as its ground state. In some models \cite{pouranvari,vitagliano2010, prb}, the presence of disorder in the couplings has been shown to give rise to strong area-law violations (more than $\ln L$-kind). Thorough characterisation of a ground state that strongly violates entanglement entropy requires the study of large systems which are computationally intractable. Furthermore, the study of such systems with numerical schemes such as density matrix renormalization group (DMRG)~\cite{schollwock2005density,schollwock2011density}, time
evolving block decimation (TEBD)~ \cite{vidal2004tebd} and other matrix product states based techniques is not feasible as they crucially depend on the area law. This leads to the motivation for simulating such systems in ultra-cold setups. 

Inspired by~\cite{prb}, we want to propose a physically realizable long range spin model with disorder in the spin-spin couplings that shows strong area-law violation of the entanglement. Quantum spin models are regularly realized in ultra-cold systems using Rydberg atoms~\cite{Robert,Schauss1455,Weimer1,Gil,Labuhn,RM}, polar molecules~\cite{Alexey,Manmana,Yan} and trapped ions \cite{molmer,Porras,Lee,Foss,Arrazola,Kim,Britton,Islam583,zoller2014}. The advantage of using ions is that they are cooled and trapped very efficiently. This allows us to manipulate the collective vibrational modes along with the internal states with great precision which is key to fine tuning the spin-spin couplings. Although most of the earlier works with trapped ions were used to simulate spin models with long range interactions without disorder, there have been some involving disorder in the potential~\cite{Smith1,Trautmann}. In this work, we consider a system of trapped ions to model our long range spin model with disorder in the spin couplings.     

In this article, we study the one dimensional long-range quantum $XY$ spin glass which contains disorder in its couplings. Using exact diagonalization, we calculate the von Neumann entanglement entropy (VNEE) for its ground state and demonstrate strong area-law violation for 14 spins. Typically, one exactly solves the 1D spin chain with nearest-neighbor spin interactions by mapping it into the free fermionic chain via the Jordan-Wigner transformation. This free fermionic mapping is not useful anymore for spin chains with (disordered) long range interactions like in our quantum $XY$ spin glass Hamiltonian because this only manages to convert one difficult problem into another difficult problem. We compare the VNEE of our model with the spinless fermionic model with disorder which is known to reliably exhibit strong area-law violation for large systems~\cite{pouranvari,prb}. We find that the VNEE of quantum $XY$ spin glass is more sensitive to the sign of the coupling than the spinless fermionic model. Also interestingly, we find strong area law violation even for short-range interactions for the quantum $XY$ spin glass system in sharp contrast to the spinless fermionic model with disorder. In the final section, we investigate the possibility of realizing the quantum $XY$ spin glass in real ion trap experiments. 

\section{Entanglement entropy for quantum $XY$ spin glass model}
The VNEE between subsystem A and the rest of the system B is defined as  
\begin{equation}
S^A =-Tr[\rho_A \ln(\rho_A)],
\label{vnee}
\end{equation}
where $\rho_A$ is the reduced density matrix of the subsystem A. VNEE
can also be described as the first order Renyi entanglement
entropy~\cite{eisert2010,amico2008entanglement,horodecki2009quantum}. In
this context, second-order Renyi entropy has been measured recently in
experiments based on ultracold atoms~\cite{Islam583,greiner2016}. The
long-range quantum $XY$ spin glass in
1D~\cite{cesare_SG,cesare2_SG,albrecht_SG} for $N$ spins is given by
\begin{align}\label{xymodel}
H_{\rm SG} &= \sum\limits_{i<j}^{N} J_{ij} (\sigma_i^+\sigma_{j}^- + {\rm h.c.}), \\
{\rm where}~J_{ij} &= J/r_{ij}^\alpha. \label{J}
\end{align}
Here $\sigma_i^+$ and $\sigma_i^-$ are the spin raising and lowering
operators respectively, acting on the $i\textsuperscript{th}$
spin. The coupling between $i\textsuperscript{th}$ and
$j\textsuperscript{th}$ spins is given by $J_{ij}$ where $J$ is chosen
from a uniform distribution of random numbers $[-V, V]$ and
$V$ is an arbitrary positive constant. This implies that between
any two spins, there can be either an anti-ferromagnetic or
ferromagnetic coupling, chosen randomly as the sign of $J_{ij}$
changes randomly. $r_{ij} = |i-j|$ is the distance between sites $i$ and $j$ in a chain and
$\alpha$ is the long-range parameter. We assume open boundary conditions throughout.
\begin{figure}
	\includegraphics[width=0.45\columnwidth]{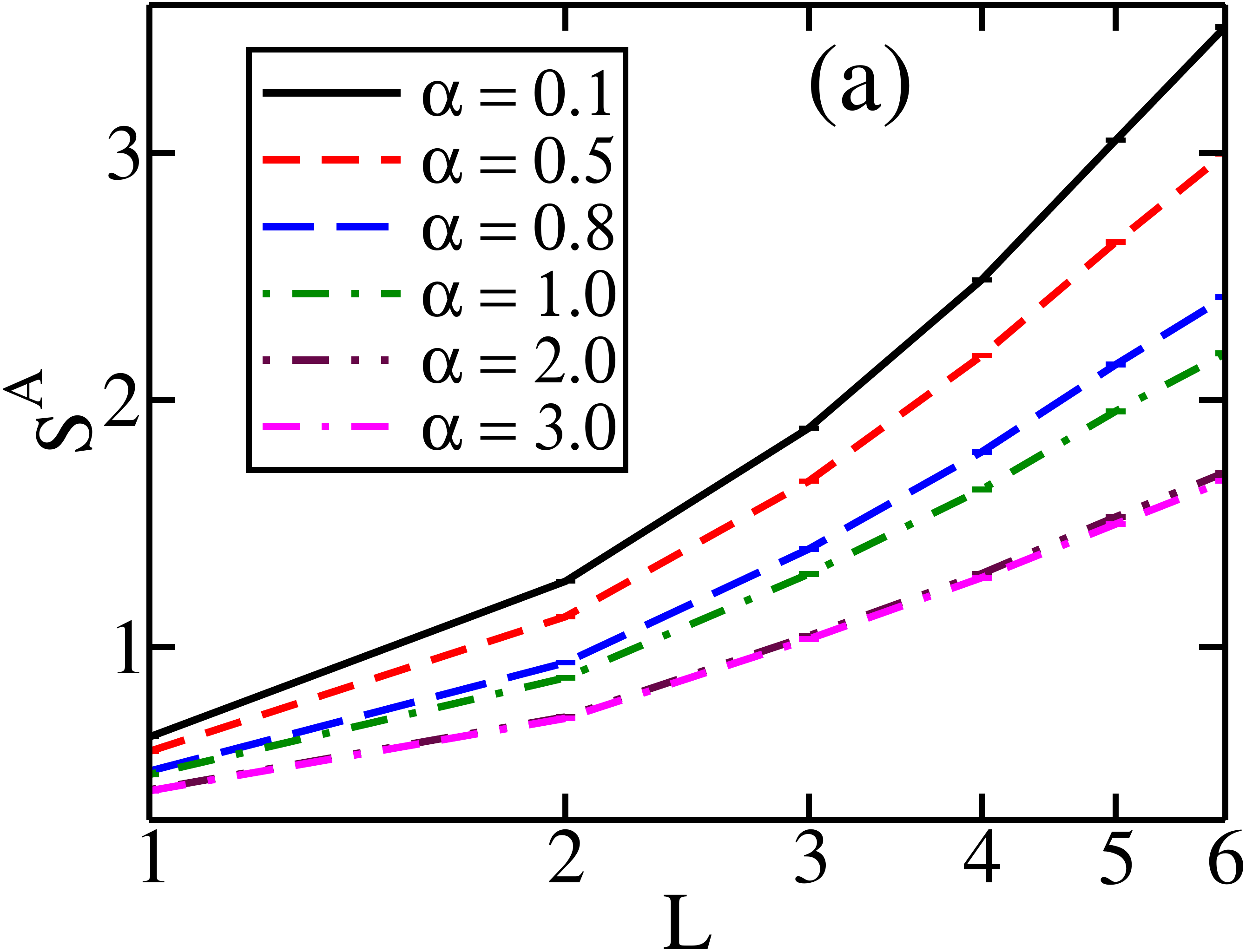}
	\includegraphics[width=0.45\columnwidth]{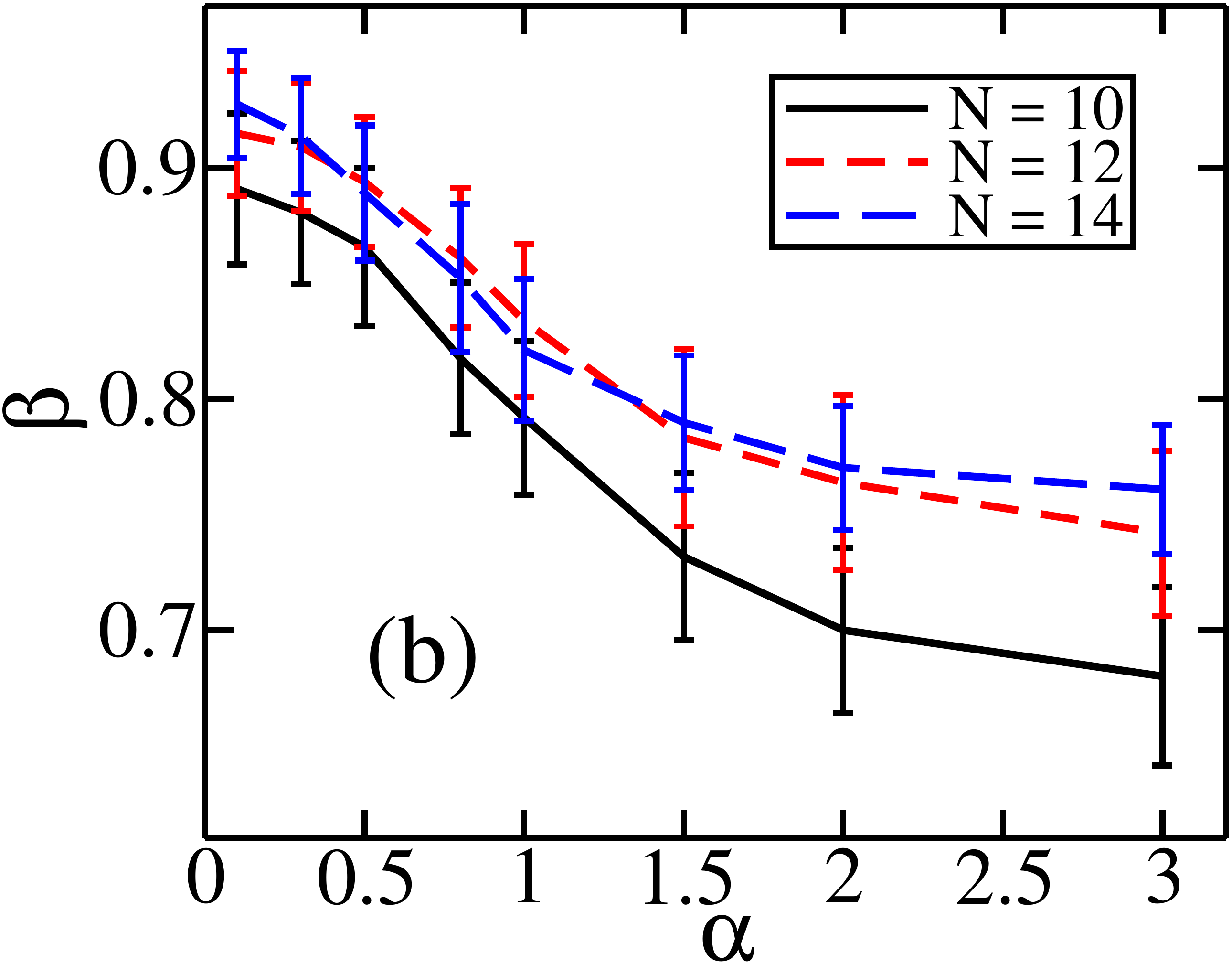}
	\caption{Panels (a) and (b) depict the characteristics of the entanglement entropy for the $1$D quantum $XY$ spin glass. For all the plots, we have taken $100$ disorder realizations. (a) Scaling of the entanglement entropy $S^A$ with the subsystem size $L$ for increasing values of $\alpha$ in the one-dimensional long-range quantum $XY$ spin glass of $N=14$ spins. The $x$-axis is in logarithmic scale. (b) The exponent $\beta$ obtained from scaling of the entanglement entropy with the subsystem size $(S^A \propto L^{\beta})$ as a function of $\alpha$ for increasing system sizes $N=10,12,$ and $14$ repectively. For all the plots $V=1$.}
	\label{combo}
\end{figure} 

In order to calculate the VNEE for Eq.(\ref{xymodel}), we first diagonalize the Hamiltonian exactly in the basis of states that correspond to an equal number of spins in the up state and in the down state. We identify the many-body ground state $\ket{\Psi_g}$ as the eigenstate corresponding to the lowest eigenvalue of the Hamiltonian and express the full density matrix as $\rho=\ket{\Psi_g}\bra{\Psi_g}$. We
partition our system of $N$ spins into subsystem A of $L$ spins and subsystem B of $N-L$ spins. The eigenvalues of the reduced density matrix $\rho_A=Tr_B[\rho]$ are obtained and denoted as $\zeta_k$ (appendices for details). Using Eq.~\ref{vnee}, the VNEE is then given by 
\begin{eqnarray}
S^A=-\sum\limits_{k=1}^{2^L} \zeta_k \ln \zeta_k.
\end{eqnarray}
Our exact numerical calculations for a system size of $N=14$ reveals power-law violation of area-law, which is depicted in Fig.~\ref{combo}(a) as a function of the parameter $\alpha$. The power-law exponents can be extracted by curve fitting, which is shown in Fig.~\ref{combo}(b) for increasing system sizes $N=10,12$ and $14$ respectively, which indicates that the power-law violation is becoming more robust with increasing system sizes. Also we have checked that the power law violation of VNEE is robust against the increasing number of disorder realizations.
\begin{figure}
	\includegraphics[width=0.45\columnwidth]{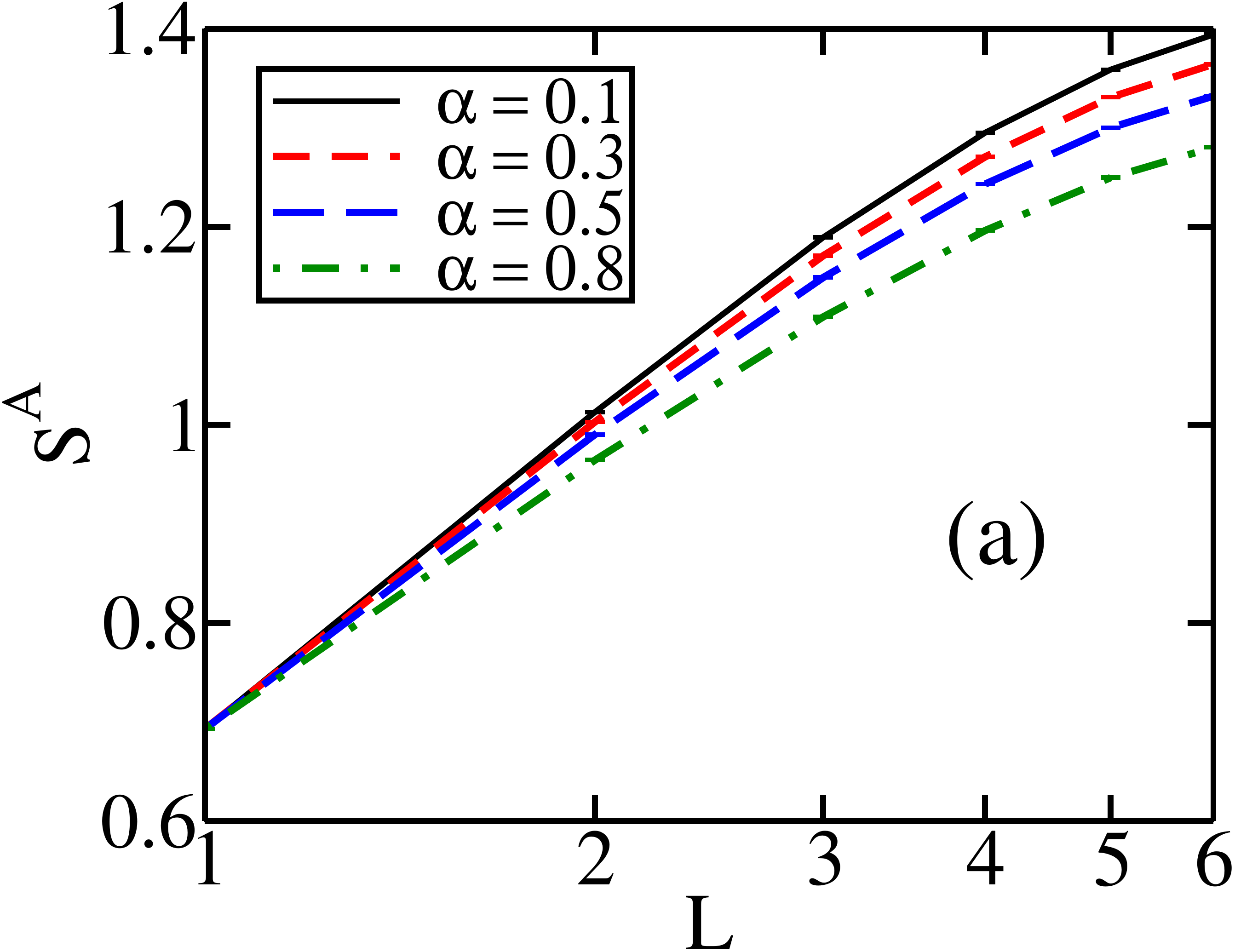}
	\includegraphics[width=0.45\columnwidth]{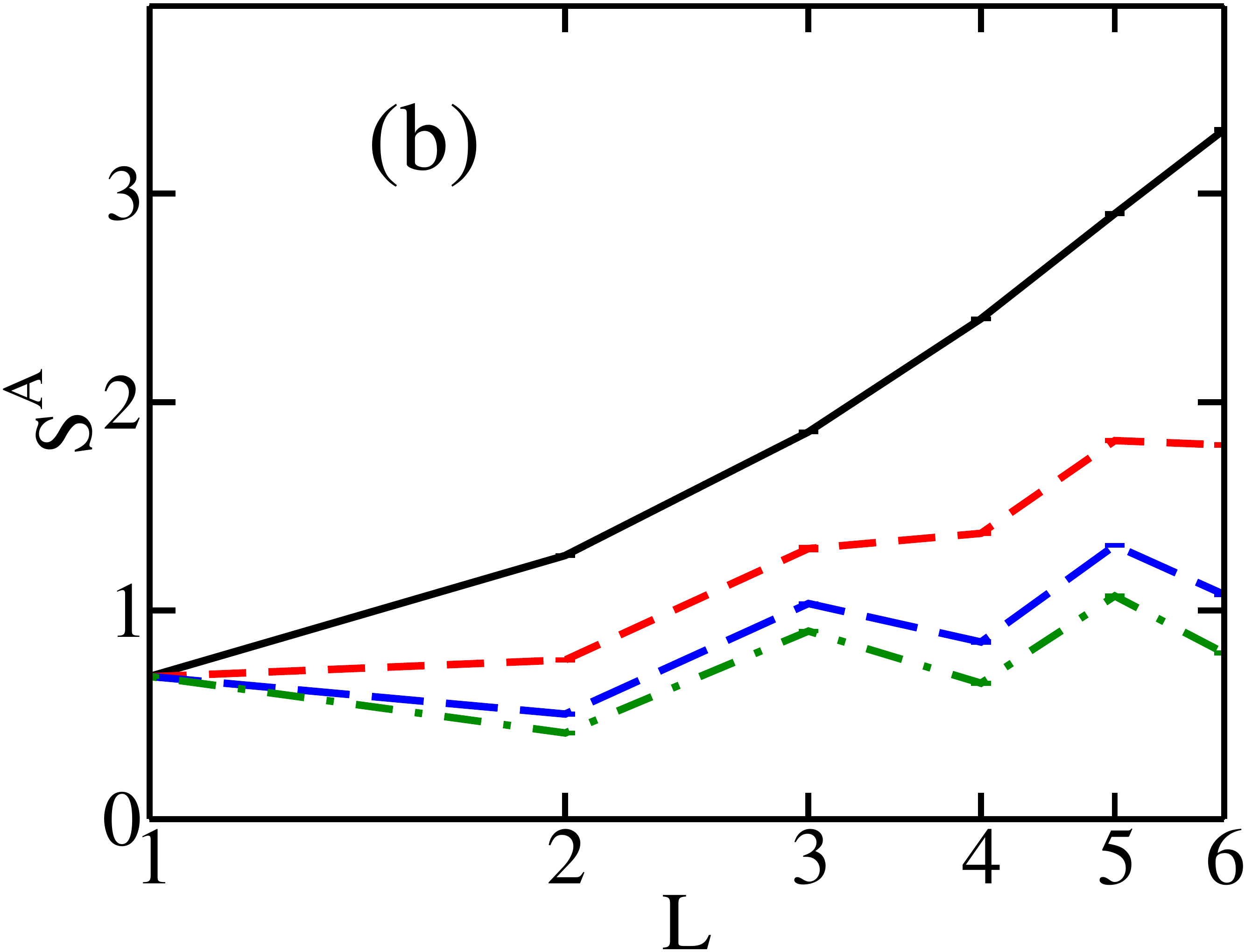}
	\caption{(a)-(b) We compare entanglement properties of long range spin chains where the $J$'s of the couplings $J_{ij}$ are chosen from the distributions of negative random numbers $[-V,0]$ and positive random numbers $[0,V]$ respectively for $N=14$ spins with open boundaries. For all the plots, we have taken $100$ disorder realizations. (a) Scaling of the entanglement entropy $S^A$ with the subsystem size $L$ (in logarithmic scale) for increasing values of $\alpha$ for the spin chain with negative random couplings. (b) The entanglement entropy $S^A$ vs. the subsystem size $L$ (in logscale) curves for the same values of $\alpha$ for the spin chain with negative random couplings. For all the plots $V=1$.}
	\label{disorder}
\end{figure}	

In order to study the role of disorder, we considered
a few variant models. We found that a disordered ferromagnet
(negative random couplings), does not show stronger than $\ln L$-kind violation no matter how long-range the couplings are, as shown
in Fig.~\ref{disorder}(a). On the other hand the disordered anti-ferromagnet
(positive random couplings) does yield power-law violation of the
area law if the long-range parameter $\alpha$ is sufficiently small, just like the spin glass as is clear from
Fig.~\ref{disorder}(b).
  
\section{Entanglement entropy for spinless fermonic model}
On the other hand, we are able to calculate the VNEE for a very large system for the spinless fermionic model which is given as
\begin{equation}
H_{F} = \sum_{i<j}^{N} J_{ij} (c_i^\dagger c_j + {\rm h.c.}).
\end{equation}
Here $c_i^\dagger$ ($c_i$) is the fermionic creation (annihilation) operator at site $i$. $J_{ij}$ is defined similarly as in Eq.(\ref{J}). One  can calculate VNEE for large systems for a noninteracting spinless fermionic model because the entanglement entropy is obtained from the eigenvalues of the correlation matrix $C_{ij}=\langle c_i^\dagger c_j\rangle$, where $i,j \in A$. This is numerically much easier to calculate as compared to the calculation of the exact reduced density matrix  (for further details see~\cite{peschel2003calculation,peschel2012special,prb} and appendics). The reason is that the ground state has a Slater determinant structure to which the  Wick's theorem may be applied. Thus the two-point correlation is sufficient to obtain all the higher order correlations.
\begin{figure}
	\includegraphics[width=0.45\columnwidth]{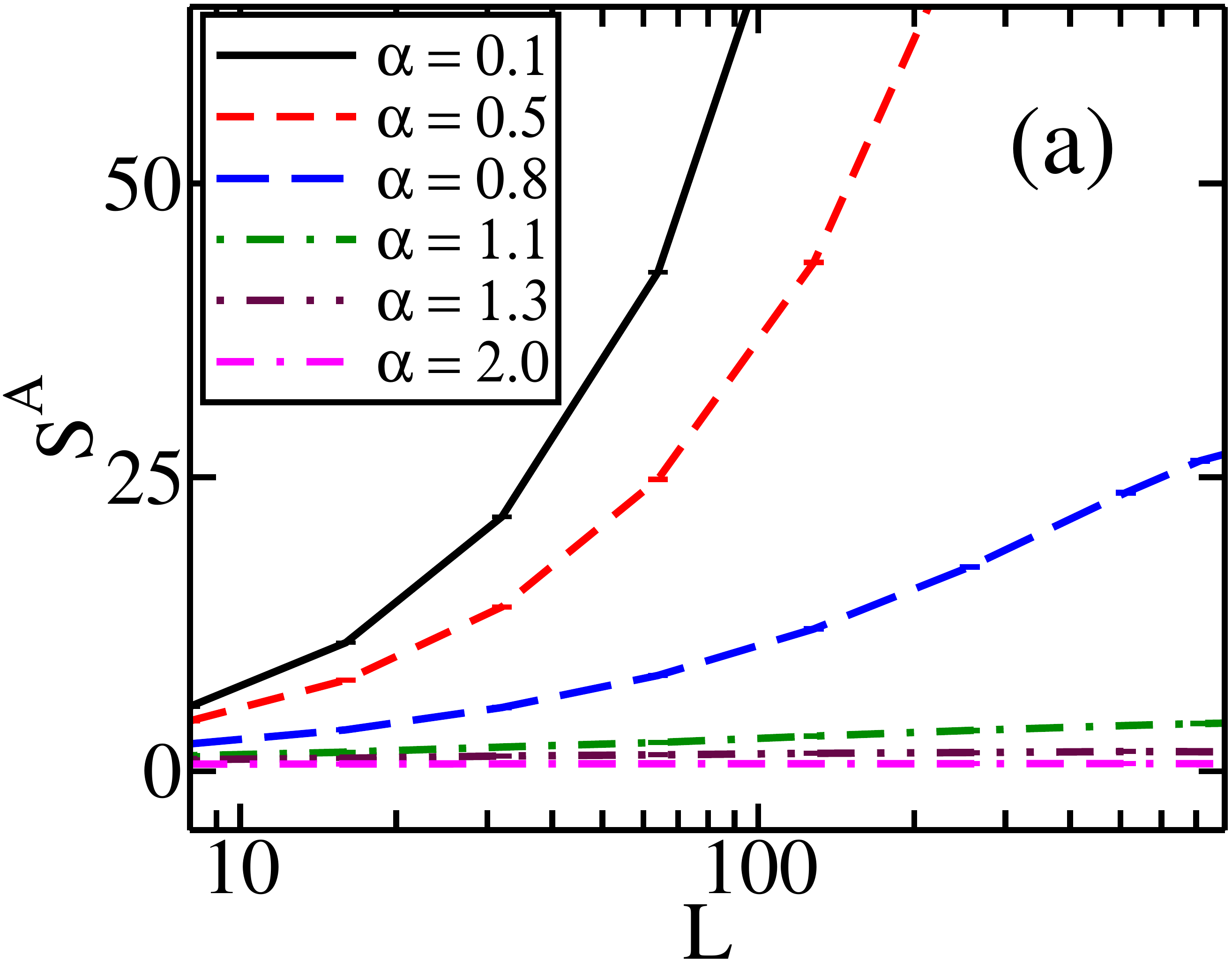}
	\includegraphics[width=0.45\columnwidth]{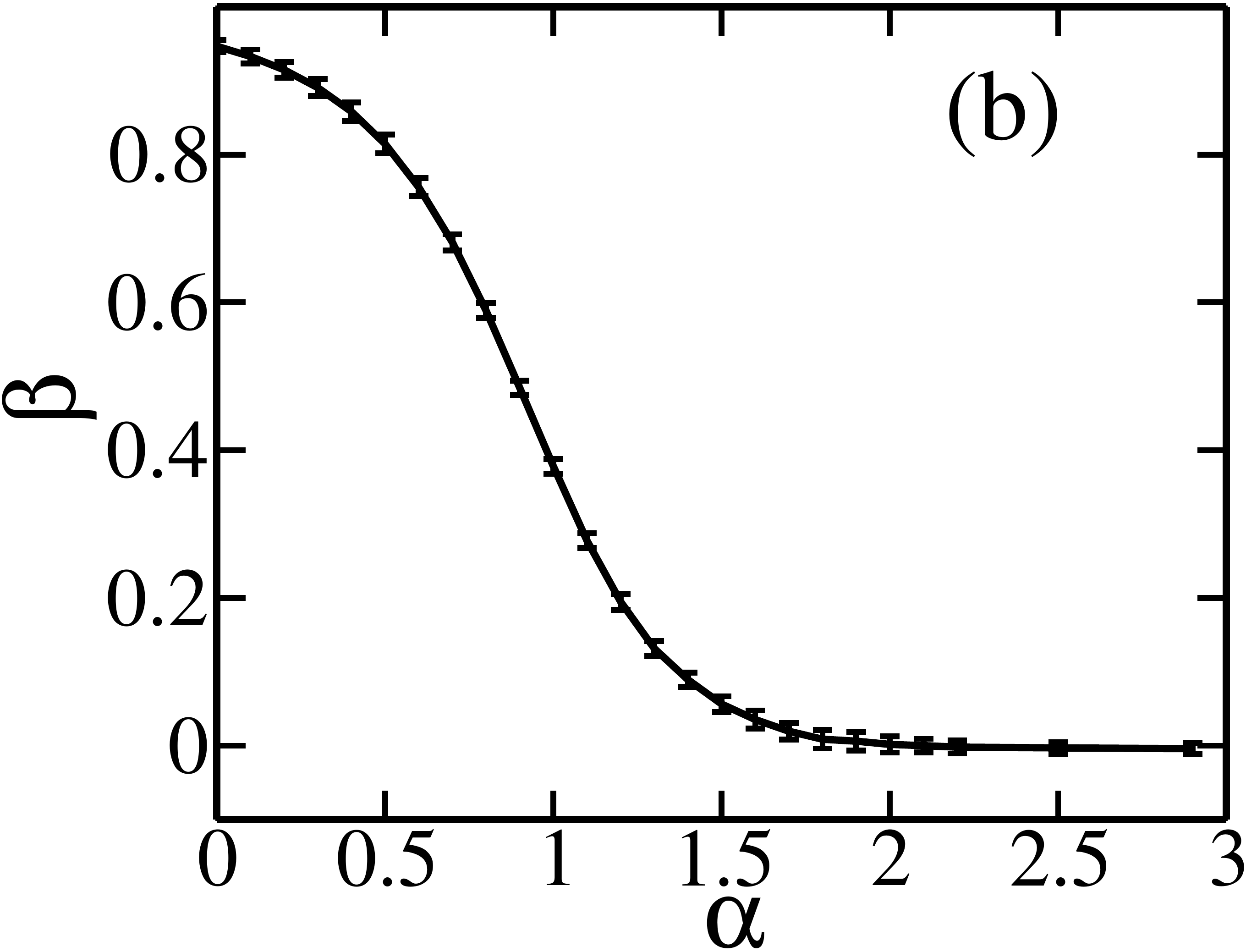}
	\caption{Panels (a)-(b) show the entanglement properties of the spinless fermionic model with $N=2048$ sites with half-filling. For all the plots, we have taken $100$ disorder realizations. (a) Scaling of the entanglement entropy $S^A$ with the subsystem size $L$ (in logarithmic scale) for increasing values of $\alpha$ in the fermionic  model. (b) The exponent $\beta$ obtained from $S^A\propto L^{\beta}$ scaling as a function of $\alpha$ for the fermionic model. For all the plots $V=1$.}
	\label{combo2}
\end{figure} 
The VNEE for the noninteracting fermionic system is
\begin{eqnarray}
S^A=-\sum\limits_{k}[\lambda_k \ln \lambda_k + (1-\lambda_k) \ln(1-\lambda_k)],
\label{vnee_f}
\end{eqnarray}
where $\lambda_k$ are the eigenvalues of $C_{ij}$. This fermionic
system shows a strong (power law type) area law violation of
entanglement entropy as can be seen in Fig.~\ref{combo2}(a) \cite{prb}. The
dependences of the power law exponents $\beta$ on the long-range
parameter $\alpha$ are shown in Fig.~\ref{combo2}(b). The exponent
$\beta$ remains non-zero in the regime $0<\alpha\leq1$ and then it
slowly vanishes as the long-range parameter $\alpha$
increases. 

Interestingly, when compared to the exponent obtained for
the quantum spin glass system, we find that it does not vanish even
for large $\alpha$. This means that there is an indication of strong
area law violation even in short range interactions but to confirm
this one would need to go to larger systems of the quantum spin glass
system. Yet another point of distinction between the ground state of
the spinless fermion model and the quantum spin glass is its
dependence on sign of the coupling. All the results of the fermionic
model remain unchanged even if the random coupling $J$ is chosen
randomly from the negative distribution $[-1,0]$ or the positive distribution $[0,1]$ whereas the spin model gives different results for small values of $\alpha$ in cases of the negative and positive random couplings as previously discussed. In order to get power law behavior across all the values $\alpha$ shown in Fig.~\ref{combo}(b), we need both ferromagnetic as well as anti-ferromagnetic couplings. These observations suggest that the underlying frustration in these models has a crucial role in the strong area law violation.

\section{Quantum simulation of long range disordered spin models}
\begin{figure}[t!]
	\includegraphics[width=1.0\columnwidth]{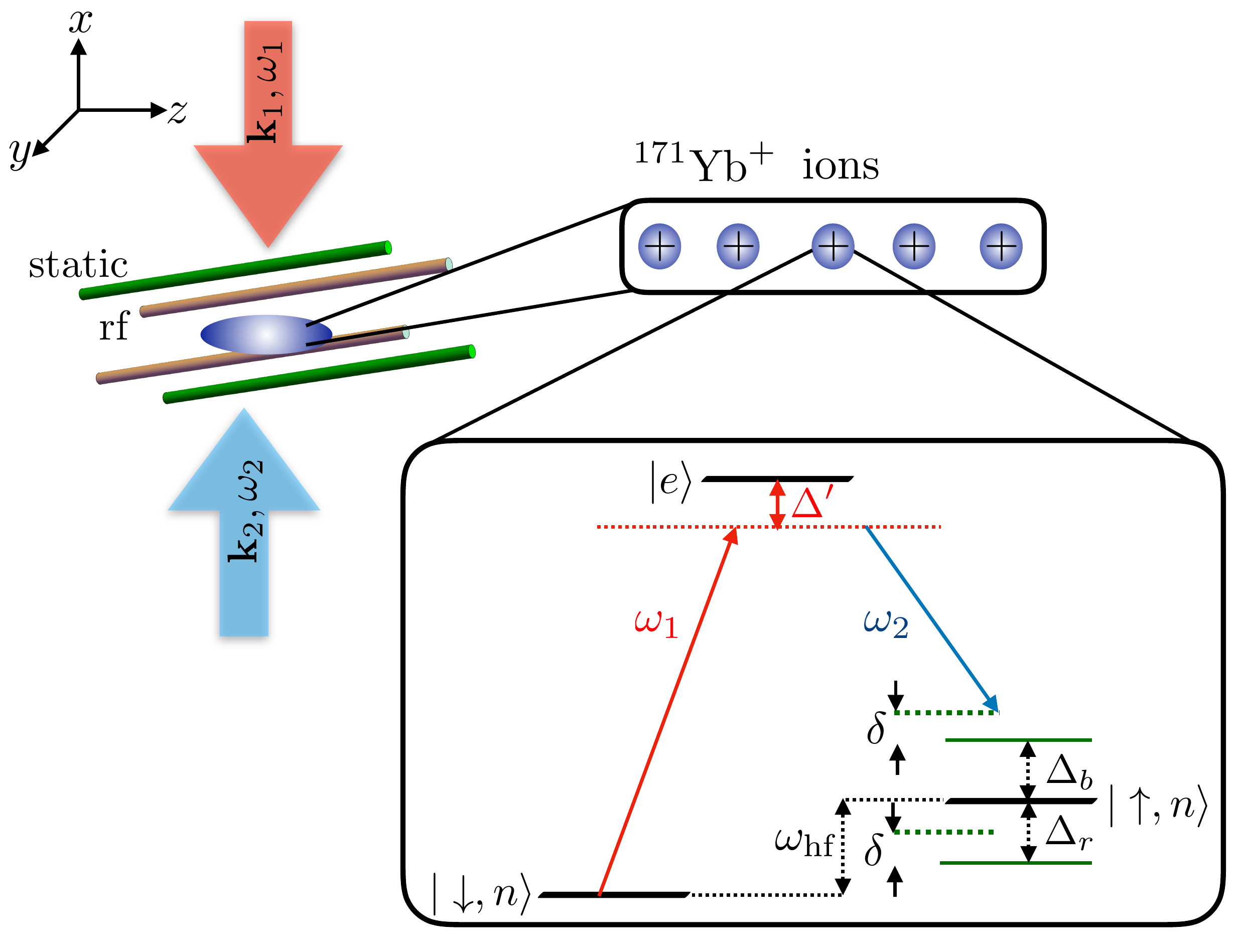}
	\caption{The figure schematically shows the setup for realising the disordered long range spin model. An array of $^{171}$Yb\textsuperscript{+} ions are trapped linearly along the $z$ axis in a Paul trap whose relevant hyper-fine states form the spin-1/2 system. The two internal states (as well as motional states)($\ket{\downarrow,n}$, $\ket{\uparrow,n}$) are coupled via the excited state $|e\rangle$ with the help of a pair of Raman lasers (show in red and blue arrows) applied along the $x$ axis with laser frequencies $\omega_{1,2}$ and wave vectors $k_{1,2}$. The pair of lasers are effectively detuned from the inter-mediate excited state $|e\rangle$ given by $\Delta'$ and have a bichromatic beat frequency ($\omega_L = \omega_2 - \omega_1$) that is simultaneously blue ($\Delta_b = \omega_L - \omega_{\rm hf} + \nu_x$) and red ($\Delta_r = \omega_L - \omega_{\rm hf} - \nu_x$) detuned from the center-of-mass trapping frequency $\nu_x$ along the transverse axis~\cite{molmer,Porras,Lee,Foss,Arrazola,Kim,Britton,Islam583}. Additionally, we have an asymmetric shift $\delta$ to the detunings which realizes the effective spin model given in Eq.(\ref{xymodel}).}

	\label{setupfig2}
\end{figure}

Next we implement the long range disordered Hamiltonian in Eq.(\ref{xymodel}) using trapped ions. There exist plenty of proposals involving trapped ions realizing various spin models~\cite{molmer,Porras,Lee,Foss,Arrazola,Kim,Britton,Islam583} including disordered potentials~\cite{Smith1,Trautmann}. However, in this work we focus on realising a spin model with disorder in its spin couplings. As shown in Fig.~\ref{setupfig2}, our setup consists of a one-dimensional chain of $^{171}$Yb\textsuperscript{+} ions aligned along the $z$-direction in a standard linear Paul trap \cite{James1998}. The effective spin-1/2 system constitutes of two hyperfine states, $^2S_{1/2}\ket{F=0,m_F=0}$ and $^2S_{1/2}\ket{F=1,m_F=0}$, with an energy splitting of $\omega_{\rm hf} = 12.64$ GHz (we assume $\hbar=1$). Using a pair of counter-propagating Raman lasers with wave vectors $\mathbf{k}_{i=1,2}$ and laser frequencies $\omega_{i=1,2}$ along the transverse $x$-axis, the two internal states are coupled via an intermediate excited state $|e\rangle$ which in our case is the $^2P_{1/2}\ket{F=0,m_F=0}$ state. The pair of lasers are far detuned (by $\Delta'$) from this excited state (see Fig.~\ref{setupfig2}) where $\Delta'$ is in the order of few hundred GHz. We define the relative wave vector as $\mathbf{k}_L = \mathbf{k}_1 - \mathbf{k}_2$ and the laser  beat frequency as $\omega_L = \omega_2-\omega_1$. The Hamiltonian for the effective optical coupling of the internal states (including motional states) in the interaction picture after applying the rotating wave approximation (RWA), is $\hat{H} = \frac{1}{2} \sum^N_{i=1} \Omega_i (\sigma^+_i e^{i( \mathbf{k}_L\cdot\hat{x}_i - \Delta t + \phi)} + {\rm h.c.})$, where $\Omega_i$ is the Rabi frequency for the $i$th ion; $\Delta = \omega_L - \omega_{\rm hf}$ is the detuning and $\phi$ is the relative phase.

The ion motional dynamics is a result of the confining linear trap and the ion-ion repulsion. Ions oscillate about their equilibrium positions giving rise to normal modes of oscillation which are classified as axial and transverse modes. We couple to the transverse modes of motion along the x-axis only, matching the direction of propagation of our Raman lasers. The Hamiltonian for motional dynamics is $\hat{H}^{\rm m}  = \sum_m \nu_m a^{\dagger}_m a_m$ where $\nu_{m}$'s are the transverse mode frequencies and $\hat{a}^{\dagger}_{m}$ and $a_{m}$ are the motional mode operators. We work in the Lamb-Dicke regime, $|\eta_{im}|\sqrt{\langle \hat{a}^{\dagger}_m \hat{a}_m \rangle} \ll 1$ where $\eta_{im} = k_L b_{i,m}\sqrt{\hbar/2M\nu_m}$ is the Lamb-Dicke parameter and $b_{i,m}$ is the transverse normal mode transformation matrix for the $i$th ion in the $m$th mode \cite{Kim2}. 

Similar to previous proposals for generating spin models \cite{molmer,Porras,Lee,Foss,Arrazola,Kim,Britton,Islam583}, our beat frequency $\omega_L$ is simultaneously detuned to red and blue sidebands of the highest normal mode frequency  which in our case is $\nu_x$. We define $\Delta_b = \Delta+\nu_x$ and  $\Delta_r= \Delta-\nu_x$. For symmetrical detuning, $\Delta_s = \Delta_b = -\Delta_r$, we obtain the well known Ising model \cite{molmer,Islam583}. However, by adding a small asymmetry $\delta$ to the effective detuning $\Delta_s$ (see appendices), we get the following effective spin Hamiltonian
\begin{align}
H_{\rm spin} &= \sum_{ij} J_{ij} \left(\sigma^-_i  \sigma^+_j  + \sigma^-_i  \sigma^-_j  e^{2i\delta t} + {\rm h.c.} \right) + \sum_{i} B_i \sigma^z_i \label{l1}\\
&\simeq \sum_{ij} J_{ij} \left(\sigma^+_i \sigma^-_j +  {\rm h.c.}\right) .\label{l2}
\end{align}  
Here we have adiabatically eliminated the motional modes by assuming $|\Delta^{\pm}_m| \gg |\eta_{im} \Omega_i|$ where $\Delta^{\pm}_m = (\Delta_s \pm \nu_m) $. The spin coupling is given as 
\begin{equation}\label{Jij}
J_{ij} =  \Omega_i \Omega_j  \sum_{m} \eta_{i,m}  \eta_{j,m}\left[ \frac{\nu_m}{\Delta^2_s-\nu^2_m}   \right]  \simeq \frac{J}{|z_i-z_j|^\alpha},
\end{equation}
where $z_{i}$ is the equilibrium position of the $i$th ion. The explicit expression for $B_i$ is given in the appendix. Eq.(\ref{l2}) is a good approximation to Eq.(\ref{l1}) provided 
\begin{equation}
|\Delta_s | \gg |\delta| \gg |J_{ij}|~\text{for all}~i,j. \\
\end{equation}
For our numerics, we consider a linear Paul trap with an axial trap frequency $\nu_z/(2\pi) = 1.0$ MHz and transverse trapping frequencies, $\nu_y/(2\pi) \simeq \nu_x/(2\pi) = 3.95$ MHz. The Rabi frequency is kept constant at $\Omega_i/2\pi = \Omega_j/2\pi = 50$ kHz and the wave numbers are  $k_1 = k_2 = 2\pi/369.75$ nm$^{-1}$. Thus for $^{171}$Yb\textsuperscript{+} ions we get $\eta_{im} \simeq 0.01 -0.03$ giving $|\eta_{im} \Omega_i/2\pi| \simeq 500$ Hz. We have $\Delta_s/2\pi = 3.5-4.5$ MHz and $\delta/2\pi \simeq 100$ kHz, for which the above conditions are easily satisfied.
\begin{figure}[t!]
	\centering
	\includegraphics[width=1.0\columnwidth]{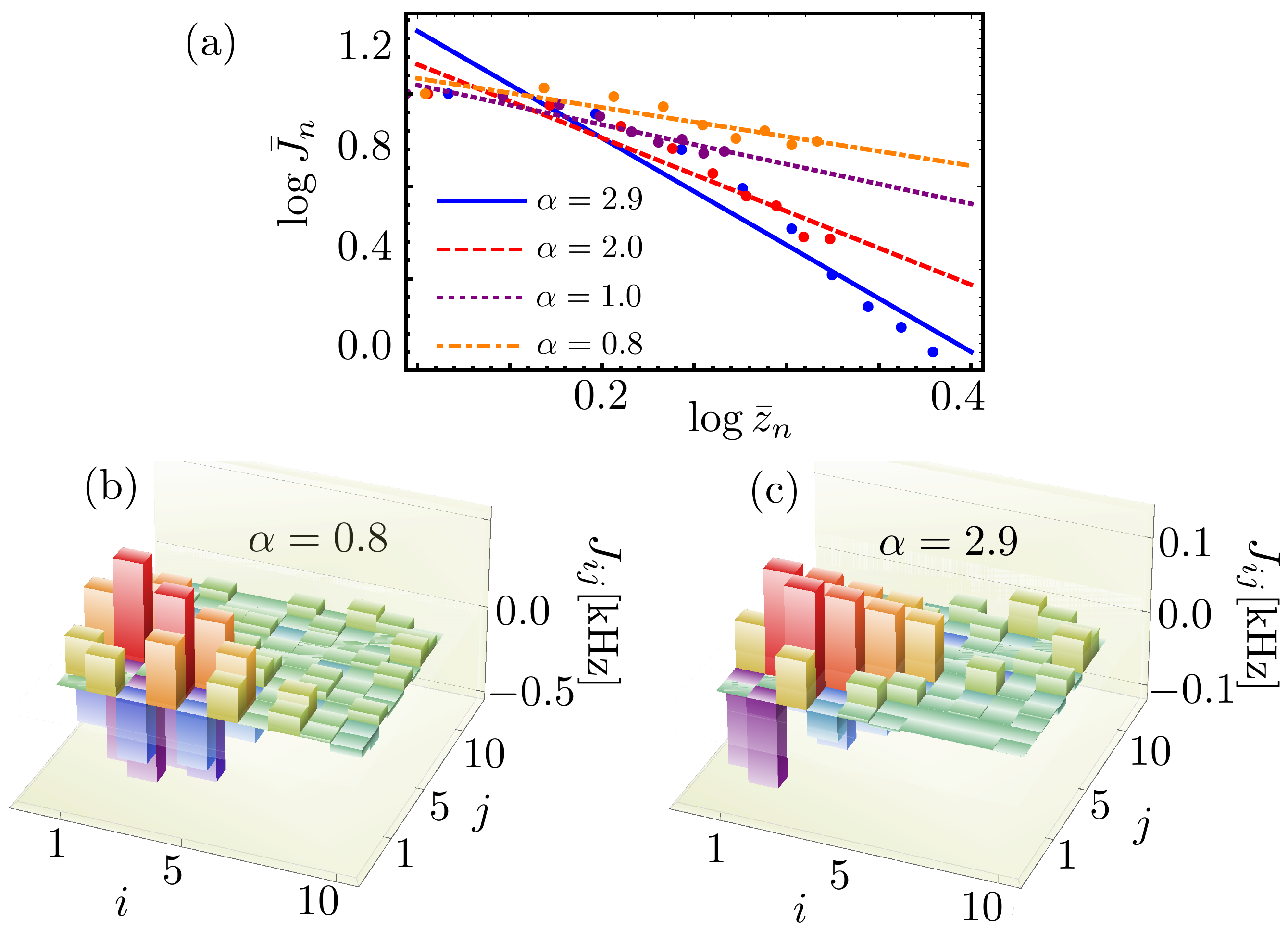}
	\caption{The figure characterizes the long range behavior as well as the disorder in the spin-spin couplings for 10 ions in 1D. (a) Plotting the normalized couplings as a function of the spin-spin separation averaged over the total spins. Here $\bar{J}_n = \sum_i |J_{i,i+n}|/(|J_{i,i+1}|(N-n))$ and $\bar{z}_n = \sum_i |z_{i+n}-z_i|/(N-n)$, where $n=1,2,\hdots$ represent the proximity of the coupling ion with respect to the $i$th ion. $z_{i}$'s are equilibrium positions of the respective ions. Couplings with different power law decays are obtained by selecting different detunings ranging from 3.9 MHz to 4.1 MHz. (b-c) We show the randomness (in sign and magnitude) of the actual couplings $J_{ij}$ for two specific values of $\alpha=0.8,2.9$.}
	\label{PlotdisorderJ_d}
\end{figure}	

Usually the spin models are generated by choosing the detuning far away from the entire bunch of mode frequencies to obtain spin couplings of the same sign. However by tuning close to the center of mass frequency along the $x$ axis, we introduce disorder in a controlled manner. To implement disordered spin interactions, we choose the bichromatic beat frequency to be slightly below and above $\nu_x$ which gives the random spin couplings that we need for our model. A spatially dependent random intensity profile constructed using spatial light modulation methods~\cite{Nogrette} or speckled intensity techniques~\cite{McGehee} can also be used to create disordered spin couplings. To obtain different power laws in $J_{ij}$ as a function of distance, we scan the detuning $\Delta_s$ over a certain range. Fig.~\ref{PlotdisorderJ_d} characterizes the power law behaviour and the disorder in our spin couplings given by Eq.(\ref{Jij}). The power law scaling in Fig.~\ref{PlotdisorderJ_d}(a) is obtained after averaging over the couplings over different ions. The power law scaling over distance in Eq.(\ref{Jij}) is usually an empirical approximation and our fittings are limited by finite size effects. Particularly for small system sizes, the power law scaling is highly sensitive to ion equilibrium positions. A pair of ions near the center of the chain are closer to each other as opposed to those at the edge. The typical inter-ionic distances for our chosen trap frequencies ranges from $1.4$ $\mu$m to $2.0$ $\mu$m. Large number of ions in a linear Paul trap can lead to unstable transverse motion and unwanted zig-zag patterns \cite{Schiffer}. Including a larger number of ions requires having lower values of $\epsilon = \nu_z/\nu_x$ which are determined by experimental constraints. The scenario involving a two dimensional Penning trap is more relaxed in this regard and good power law fits are obtained for hundreds of ions \cite{Britton} but have yet to be shown for disordered couplings. Fig.~\ref{PlotdisorderJ_d}(b) and (c) depict the disorder in the couplings for different values of $\alpha$. One clearly sees the random fluctuation in sign and magnitude of $J_{ij}$ for different values of $i,j$. Although, one can couple to axial modes as well which are typically at a much lower frequency and are more widely separated compared to the transverse modes, our calculations seem to suggest that there is no particular advantage in choosing either type of mode for small systems.

\section{Conclusion}
To summarize, we have studied the entanglement properties of the ground state of the long-range one dimensional quantum $XY$ spin glass. We find that it exhibits a strong area-law violation, which in contrast with the disordered long-range fermionic hopping model, extends into the short-range regime.  Such strong area-law violation spells doom for numerical techniques to be effective beyond extremely small system sizes. This motivates the implementation of a quantum simulation approach to access such models.  We thus consider trapped ions to realize our spin Hamiltonian which can be generalized to larger systems in future experiments. Our work also indicates the possibility of achieving strong area-law violating ground states using disordered long ranged couplings. Quantum spin models with disordered long range couplings have not been explored much so far either theoretically or experimentally. We hope that our work will help boost more interseting research engaging such spin models. 

We are grateful to the High Performance Computing (HPC) facility at IISER Bhopal, where large-scale calculations in this project were run. N.R is grateful to University Grants Commission (UGC), India for support via his Ph.D fellowship. R.M would like to acknowledge J. Lishman for his useful comments and the Max-Planck society for funding under the MPG-IISER partner group program and . A.S acknowledges SERB for the startup grant (File Number: YSS/2015/001696).

\bibliography{refs}

\clearpage

\subsection*{Appendix A: Methodology to calculate entanglement entropy for quantum spin glass}\label{appA}
The spin glass Hamiltonian is given by
\begin{equation}
H_{SG} =  \sum\limits_{i<j=1}^{N} J_{ij} (\sigma_{i}^{+}\sigma_{j}^{-} + h.c.), 
\end{equation}
where $N$ is the total number of spins with spin-$\frac{1}{2}$. The Hamiltonian matrix is  written in the $\sigma_z$ basis ($\ket{\uparrow}$ and $\ket{\downarrow}$) where total magnetization is zero i.e. number of up spins is equal to the number of down spins. $H_{SG}$ is exactly diagonalized in the basis and we obtain the normalized many-body ground state $\ket{\Psi_g}=\sum_{l} \psi_l \ket{l}$. The corresponding density matrix is given by $\rho=\ket{\Psi_g}\bra{\Psi_g}$, which can be written as
\begin{eqnarray}
\rho=\sum\limits_{l,l^\prime=1}^{\textsuperscript{N}C_{N/2}} \psi_l \psi_{l^\prime}^* \ket{l}\bra{l^\prime}. 
\end{eqnarray}
We consider the first $L$ (where $L\leq N/2$) spins as the subsystem A and the remaining $N-L$ spins as the subsystem B. Each element in the basis then can be written as $\ket{l}=\ket{l_A}\ket{l_B}$, where $\ket{l_A}$ and $\ket{l_B}$ are the orthonormal sets in subsystem A and B respectively. The reduced density matrix defined as $\rho_A=Tr_B[\rho]$, can be written as 
\begin{eqnarray}
\rho_A=\sum\limits_{l_A,l_A^\prime} \rho_A^{l_A,l_A^\prime} \ket{l_A}\bra{l_A^\prime},
\end{eqnarray}
where $\rho_A^{l_A,l_A^\prime}$ is the $(l_A,l_A^\prime)\textsuperscript{th}$ element of matrix $\rho_A$. One then diagonalizes the matrix $\rho_A$ exactly and the von Neumann entanglement entropy $S^A=-Tr[\rho_A \ln \rho_A]$ is computed with
\begin{eqnarray}
S^A=-\sum\limits_{k=1}^{2^L} \zeta_k \ln \zeta_k.
\end{eqnarray}
where $\zeta_k$'s are the eigenvalues of $\rho_A$.

\subsection*{Appendix B: Methodology to calculate entanglement entropy for spinless fermionic model}\label{appB}
The fermionic Hamiltonian is given by
\begin{equation}
H_F =  \sum\limits_{i<j=1}^{N} J_{ij} (c_{i}^{\dagger}c_{j} + h.c.) 
\end{equation}
whose diagonal form is
\begin{equation}
H_F =  \sum\limits_{k=1}^{N} \epsilon_{k} b_{k}^{\dagger}b_{k},
\end{equation} 
where $b_k=\sum\limits_{j=1}^{N}\psi_j(k) c_j$. The many-body fermionic ground state is
\begin{equation}
\ket{\Psi_0}=\prod_{k=1}^{N_p}b_k^\dagger\ket{0}
\end{equation}
Due to the Slater determinant structure of $\ket{\Psi_0}$, all higher correlations can be obtained by two point correlation $C_{ij}=\langle c_i^\dagger c_j\rangle$\cite{peschel2003calculation,peschel2009,peschel2012special}. 
The density matrix of the full system is $\rho=\ket{\Psi_0}\bra{\Psi_0}$ and the reduced density matrix of subsystem A is $\rho_{A}=Tr_{B}(\rho)$. By definition a one particle function, in this case two-point correlation in the subsystem, can be written as
\begin{equation}
C_{ij}=Tr[\rho_A c_i^\dagger c_j]=Tr[\rho c_i^\dagger c_j],
\label{correlation}
\end{equation}
where $i,j\in A$. The subsystem density matrix of an eigenstate of a quadratic fermionic Hamiltonian is thermal\cite{peschel2003calculation}
\begin{equation}
\rho_{A}=\frac{e^{-H_{A}}}{Z},
\end{equation}
where $H_{A}=\sum\limits_{i,j=1}^{L} H_{ij}^A c_{i}^{\dagger}c_{j}$ is called the
entanglement Hamiltonian, and $Z$ is obtained to satisfy the condition $Tr[\rho_{A}] = 1$. The entanglement Hamiltonian can thus be written in the diagonal form as,
\begin{equation}
H_A=\sum\limits_{k=1}^{L} h_k a_k^\dagger a_k,
\end{equation}
where $a_k=\sum_{j=1}^{L} \phi_j(k) c_j$. The reduced density matrix is then given by
\begin{equation}
\rho_A=\frac{exp[-\sum\limits_{k=1}^{L}h_k a_k^\dagger a_k]}{\prod\limits_{k=1}^{L}[1 + exp(-h_k)].}
\label{rho_A}
\end{equation}
Using Eq.\ref{rho_A}, we can write Eq.\ref{correlation} as,
\begin{equation}
C_{ij}=\sum\limits_{k=1}^{L} \phi^\ast_i(k){\phi}_j(k) \frac{1}{e^{h_k}+1}.
\end{equation}
This shows that the matrices $C$ and $H_A$ share the eigenstate $\ket{\phi_k}$ and their eigenvalues are related by
\begin{equation}
\lambda_k=\frac{1}{e^{h_k}+1},
\label{relation}
\end{equation}
where $\lambda_k$'s are eigenvalues of matrix $C$ in the subsystem.\\
The von Neumann entanglement entropy can be simplified~\cite{sharma2015landauer} using Eq.\ref{rho_A} and Eq.\ref{relation} as
\begin{equation}
S^A = - \sum\limits_{k=1}^{L} [\lambda_k \ln \lambda_k + (1-\lambda_k) \ln (1-\lambda_k)].
\end{equation}

\subsection*{Appendix C: Effective spin model using time dependent perturbation theory}\label{appC}

Specific spin-spin interactions are induced by the excitation lasers and trapping conditions. Our bichromatic beat frequency is tuned symmetrically to the red and blue sidebands simultaneously and thus we have $\Delta_s = \Delta_b = -\Delta_{r}$. In the Lamb-Dicke regime, we expand $\mathbf{k}_L\cdot\hat{x}_i$ in terms of the motional operators to obtain the well known spin dependent Hamiltonian \cite{Foss,Arrazola,Kim,Britton,Islam583}
\begin{equation}
\hat{H}  =\sum_{i,m} \Omega_i \eta_{im} \sigma^x \cos(\Delta_s t) \left[a_m e^{-i\nu_m t} + a^{\dagger}_m e^{i\nu_m t}\right] .
\end{equation}
Here we have assumed that the relative phase for the pair lasers is $-\pi/2$. As discussed in the text, we can make our detuning asymmetric by shifting the levels by $\delta$, i.e. $\Delta_s + \delta = -\Delta_r + \delta = \Delta_b + \delta$, which essentially implies that our ladder operators rotate in the manner: $\sigma^{\pm} \rightarrow \sigma^{\pm}e^{\mp i \delta t}$ and thus our Hamiltonian gets modified to 
\begin{equation}
\hat{H}  = \sum_{i,m} \Omega_i \eta_{im} \sigma^{r}_i G_m(t)
\end{equation}
where we define
\begin{align}
\sigma^r_i &= \sigma^+_i e^{-i\delta t} + \sigma^-_i e^{i\delta t} \\
G_m(t) &= a_m( e^{i\Delta^{-}_m t} + e^{-i\Delta^{+}_m t}) + a^{\dagger}_m (e^{-i\Delta^{-}_m t} + e^{i\Delta^{+}_m t} ) 
\end{align}
and $\Delta^{\pm}_m = \Delta_s \pm \nu_m$. For the parameters chosen in our article, typical values of $\Delta^{\pm}_m$ are slightly less than one MHz. Using the Magnus expansion, we can write the time evolution operator as follows
\begin{equation}\label{U}
U(t) = e^{-iH(t)}=e^{ H^{(1)}(t)+H^{(2)}(t) + \hdots},
\end{equation}
where $H^{(1)}(t) = -i \int^t_0 H(t_1) dt_1$ and $H^{(2)}(t) = -\frac{1}{2} \int^t_0 \int^{t_1}_0 [H(t_1),H(t_2)] dt_1 dt_2$. $H^{(1)}(t)$ corresponds to the spin-phonon coupling and can be ignored if we have $\delta \pm \Delta^{\pm}_m \gg |\eta_{im} \Omega_i|$. For our parameters, $\eta_{im} \simeq 0.01 -0.03$ giving $|\eta_{im} \Omega_i| \simeq 500$ Hz. The smallest value of $\Delta^{\pm}_m$ is around $880$ kHz and $\delta$ can be chosen between 10 to 100 kHz for the spin-phonon coupling to be ignored. The second order term gives our effective spin model which can be split into two parts as follows
\begin{equation}
H^{(2)}(t) = \int^{t}_0 idt_1 \left(H^{(2)}_I(t_1) + H^{(2)}_{II}(t_1)  \right),
\end{equation}
where 
\begin{align}
H^{(2)}_I(t_1) &= \frac{i}{2} \sum_{i,j} \sum_{m,m'} \eta_{i,m} \eta_{j,m'} \Omega_i \Omega_j  \nonumber \\
&\times\int^{t_1}_0  [\sigma^r_i(t_1),\sigma^r_j(t_2)]G_m(t_1)G_{m'}(t_2) dt_2 \\
H^{(2)}_{II}(t_1) &= \frac{i}{2} \sum_{i,j} \sum_{m,m'} \eta_{i,m} \eta_{j,m'} \Omega_i \Omega_j   \nonumber \\
&\times\int^{t_1}_0  [G_m(t_1), G_{m'}(t_2)]\sigma^0_i(t_1)\sigma^0_j(t_2) dt_2
\end{align}
Using the commutator relationships
\begin{align}
&[\sigma^r_i(t_1), \sigma^r_j(t_2)] = 2\sigma^z_j \delta_{ij} (e^{i\delta(t_2-t_1)} - e^{-i\delta(t_2-t_1)}), \\ \nonumber \\
&[G_m(t_1), G_{m'}(t_2)] \nonumber \\
&= 2 i \delta_{mm'} \left[ \sin (\Delta^-_m(t_1 -t_2)) + \sin (\Delta^-_m t_1 + \Delta^+_m t_2)   \right. \nonumber \\
&\left. - \sin (\Delta^+_m t_1 + \Delta^-_m t_2) - \sin (\Delta^+_m(t_1 -t_2)) \right],
\end{align}
and solving the integrals over time gives
\begin{align}
H_{\rm spin} &= \sum_{ij} J_{ij} \left(\sigma^-_i  \sigma^+_j  + \sigma^-_i  \sigma^-_j  e^{2i\delta t} + {\rm h.c.} \right) + \sum_{i} B_i \sigma^z_i
\end{align}
where the definition for the  spin-spin coupling is
\begin{align}
J_{ij} &= \Omega_i \Omega_j\sum_{m} \eta_{i,m}  \eta_{j,m}  \left[ \frac{\Delta^-_m}{(\Delta^-_m)^2 - \delta^2} - \frac{\Delta^+_m}{(\Delta^+_m)^2 - \delta^2}   \right]  \nonumber \\
&\simeq \Omega_i \Omega_j \sum_{m} \eta_{i,m}  \eta_{j,m} \left[ \frac{\nu_m}{\Delta^2_s-\nu^2_m}   \right] 
\end{align}
and the longitudinal field is
\begin{align}
B_i  = & \sum_{m} (\eta_{i,m} \Omega_i)^2 \left[  \left(\frac{2\delta}{(\Delta^{-}_m)^2 - \delta^2} \right) \right. \nonumber \\
&\left. + \left(\frac{2\delta}{(\Delta^{+}_m)^2 - \delta^2} \right)  \right] \left(a^{\dagger}_m a_m + \frac{1}{2} \right) \nonumber \\
&\simeq 4\delta \sum_{m} (\eta_{i,m} \Omega_i)^2  \left[\frac{\Delta^2_s+\nu^2_m}{(\Delta^2_s-\nu^2_m)^2} \right]\left(a^{\dagger}_m a_m + \frac{1}{2} \right).
\end{align}
The approximated expressions of $J_{ij}$ and $B_i$ are obtained by assuming $|\Delta_m^{\pm}|>> \delta$.

\end{document}